\documentclass[5p,authoryear]{elsarticle}

\usepackage{natbib}
\usepackage{amsmath}
\usepackage{graphicx}
\usepackage{verbatim}
\usepackage{rotating}
\usepackage{hyperref}

\journal{Social Networks}

\begin{document}

\begin{frontmatter}

\title{Anomaly Detection in Online Social Networks}

\author[rmit1]{David Savage\corref{cauthor}}
\cortext[cauthor]{Corresponding author}
\ead{david.savage@rmit.edu.au}

\author[rmit1]{Xiuzhen Zhang}
\author[rmit2]{Xinghuo Yu}
\author[austrac]{Pauline Chou}
\author[rmit2]{Qingmai Wang}

\address[rmit1]{School of Computer Science and Information Technology, RMIT University, GPO Box 2476, Melbourne, Victoria, 3001, Australia}
\address[rmit2]{School of Electrical and Computer Engineering, RMIT University, GPO Box 2476, Melbourne, Victoria, 3001, Australia}
\address[austrac]{Australian Transaction Reports and Analysis Centre, PO Box 13173, Law Courts, Melbourne, Victoria, 8010, Australia}

\begin{abstract}

Anomalies in online social networks can signify irregular, and often illegal behaviour. Detection of such anomalies has been used to identify malicious individuals, including spammers, sexual predators, and online fraudsters. In this paper we survey existing computational techniques for detecting anomalies in online social networks. We characterise anomalies as being either static or dynamic, and as being labelled or unlabelled, and survey methods for detecting these different types of anomalies. We suggest that the detection of anomalies in online social networks is composed of two sub-processes; the selection and calculation of network features, and the classification of observations from this feature space. In addition, this paper provides an overview of the types of problems that anomaly detection can address and identifies key areas for future research.

\end{abstract}


\end{frontmatter}


\section{Introduction}

Anomalies arise in online social networks as a consequence of particular individuals, or groups of individuals, making sudden changes in their patterns of interaction or interacting in a manner that markedly differs from their peers. The impacts of this anomalous behaviour can be observed in the resulting network structure. For example, fraudulent individuals in an online auction system may collaborate to boost their reputation. Because these individuals have a heightened level of interaction, they tend to form highly interconnected subregions within the network \citep{Pandit:2007}. In order to detect this type of behaviour, the structure of a network can be examined and compared to an assumed or derived model of normal, non-collaborative interaction. Regions of the network whose structure differs from that expected under the normal model can then be classified as anomalies (also known as outliers, exceptions, abnormalities, etc.).

In recent times, the rise of online social networks and the digitisation of many forms of communication has meant that online social networks have become an important part of social network analysis (SNA). This includes research into the detection of anomalies in social networks, and numerous methods have now been developed. This development has occurred over a wide range of problem domains, with anomaly detection being applied to the detection of important and influential network participants (e.g. \citet{Shetty:2005,Malm:2011,Cheng:2013}), clandestine organisational structures (e.g. \citet{Shetty:2005,Krebs:2002,Reid:2005}), and fraudulent and predatory activity (e.g. \citet{Phua:2010,Fire:2012,Chau:2006,Akoglu:2013,Pandit:2007}).

Since anomaly detection is coming to play an increasingly important role in SNA, the purpose of this paper is to survey existing techniques, and to outline the types of challenges that can be addressed. To our knowledge this survey represents the first attempt to examine anomaly detection with a specific focus on social networks. The contributions of this paper are as follows

\begin{itemize}
\item provide an overview of existing challenges in a range of problem domains associated with online social networks that can be addressed using anomaly detection
\item provide an overview of existing techniques for anomaly detection, and the manner in which these have been applied to social network analysis
\item explore future challenges for online social networks, and the role that anomaly detection can play
\item outline key areas where future research can improve the use of anomaly detection techniques in SNA
\end{itemize}

In drafting this review we did not set out to consider particular problem domains. Rather, we aimed to identify tools specifically designed for detection of anomalies, regardless of the particular social networks they were designed to analyse. However, as we conducted our survey we found that relevant work was predominantly published in the area of computer science, and consequently, many of the applications of anomaly detection that we encountered were focused on anomalies in online systems. Therefore, unless specifically stated otherwise, the term social network will be used throughout this paper to mean an online social network.

Within the social sciences literature, we found a number of papers focusing on the concept of network change (see for example \citet{McCulloh:2011,Arney:2013,Tambayong:2014}), which attempts to characterise the evolution of social networks. We see anomaly detection as being a subset of of change detection, as anomaly detection could be used to identify change points where an evolving social network undergoes a rapid change, however a network that evolves in a consistent fashion over an extended period of time is unlikely to be deemed anomalous. We have therefore elected to limit the scope of our review to those methods that deal specifically with anomaly detection.

\section{Related Work}

Previous reviews of anomaly detection have provided an overview of the general, non-network based problem, describing the use of various algorithms and the particular types of problems to which these algorithms are most suited \citep{Hodge:2004,Chandola:2009a,Markou:2003a,Markou:2003b}. A workshop on the detection of network based anomalies was also held at ACM 2013 \citep{Akoglu:2013b}. The most recent review of general anomaly detection \citep{Chandola:2009a}, expands on previous works to define six categories of anomaly detection techniques; classification (supervised learning), clustering, nearest neighbour, statistical, information theoretic, and spectral analysis.

As well as categorising anomaly detection techniques, previous reviews describe a number of challenges for anomaly detection, mainly associated with the problem of defining normal behaviour, particularly in the face of evolving systems, or systems where anomalies result from malicious activities \citep{Chandola:2009a,Hodge:2004}. In particular, \citet{Chandola:2009a} note that the development of general solutions to anomaly detection remains a significant challenge and that novel methods are often developed to solve specific problems, accommodating the specific requirements of these problems and the specific representation of the underlying systems. As discussed in Section \ref{Discussion}, this has also been the case for some methods focused on anomaly detection in social networks.\\

In addition to the major reviews of anomaly detection described above, other works have considered anomaly detection as part of methodological surveys for particular problem domains. For example, methods for performing anomaly detection have been discussed as part of more general reviews in areas of fraud detection \citep{Bolton:2002,Phua:2010}, network intrusion \citep{Jyothsna:2011,Gogoi:2011,Patcha:2007}, and the deployment of wireless sensor networks \citep{Zhang:2010,Janakiram:2006}. While significant overlap exists between the analysis of computer and sensor networks and social networks, there are also a number of differences that must be taken into account. In particular, social networks are typically composed of many inter-connected communities, which has important consequences for the distribution of node degree, and the transitivity of the network \citep{Newman:2003}. Moreover, anomaly detection in both sensor and computer networks is typically required to occur online in (soft) real-time, and while this constraint may also apply in some SNA scenarios, it is not typically required. In addition, anomaly detection in sensor networks generally requires algorithms that reduce network traffic and have a low computational complexity \citep{Zhang:2010}.

\section{Problem domains for the application of anomaly detection in social networks}

Anomalies in social networks are often representative of illegal and unwanted behaviour. The recent explosion of social media and online social systems, means that many social networks have become key targets for malicious individuals attempting to illegally profit from, or otherwise cause harm to, the users of these systems. 

Many users of online social systems such as Facebook, Google+, Twitter, etc. are regularly subjected to a barrage of spam and otherwise offensive material \citep{Shrivastava:2008,Fire:2012,Akoglu:2010,Hassanzadeh:2012}. Moreover, the relative anonymity and the unsupervised nature of interaction in many online systems provides a means for sexual predators to engage with young, vulnerable individuals \citep{Fire:2012}. Since the perpetrators of these behaviours often display patterns of interaction that are quite different from regular users, they can be identified through the application of anomaly detection techniques. For example, sexual predators often interact with a set of individuals who are otherwise unconnected, leading to the formation of star like structures \citep{Fire:2012}. These types of structures can be identified by examining a range of network features \citep{Akoglu:2010,Shrivastava:2008,Hassanzadeh:2012}, or through the use of trained classifiers \citep{Fire:2012}.

Online retailers and online auctions have also become a key target for malicious individuals. By subverting the reputation systems of online auction systems, fraudsters are able to masquerade as honest users, fooling buyers into paying for expensive goods that are never delivered. This process is facilitated by the use of Sybil attacks (the use of multiple fake accounts) and through collaboration between fraudulent individuals to artificially boost reputation to a point where honest buyers are willing to participate in large transactions \citep{Chau:2006,Pandit:2007}. In many online stores, opinion spam, in the form of fake product reviews, is used in an attempt to distort consumers' perceptions of product quality and to influence buyer behaviour \citep{Akoglu:2013}. Again, the malicious individuals who engage in these types of behaviour often form anomalous structures within the network, as their patterns of interaction can be quite different from regular users.
 
In addition to the social networks supported by dedicated online systems, mining of the social networks induced by mobile phone communications, financial transactions, etc. can also be used to identify illegal activities. Detection of anomalies in these types of networks have previously been used to identify organised criminal behaviour, including insurance fraud \citep{Subelj:2011}, and terrorist activities \citep{Reid:2005,Krebs:2002}. Given the highly detrimental impact of these types of behaviour, anomaly detection in social networks can be seen as an extremely important component in the growing tool-box for performing social network analysis (SNA).
 
Outside of criminal or malicious behaviour, anomaly detection has also been used to detect important and influential individuals \citep{Shetty:2005}, individuals fulfilling particular roles within a community \citep{Welser:2011}, levels of community participation \citep{Bird:2008}, and unusual patterns in email traffic \citep{Eberle:2007}.

\section{Definitions}

Anomalies are typically defined in terms of deviation from some expected behaviour. A recent review of general, non-network based anomaly detection defined anomalies as ``patterns in data that do not conform to a well defined notion of normal behaviour'' \citep{Chandola:2009a}. Another recent review defines anomalies as ``an observation (or subset of observations) which appears to be inconsistent with the remainder of that set of data'' (\citep{Barnett:1994}, cited in \citep{Hodge:2004}). We note two important aspects of these definitions.

First, the definitions given above highlight the importance of defining expected behaviour, but are somewhat imprecise in terms of exactly how the anomaly deviates from this expectation. This imprecision stems from the fact that the magnitude of any deviation will depend on the specific problem domain \citep{Chandola:2009a}.

Second, these definitions presuppose an understanding of what exactly should be observed in order to differentiate normal and anomalous behaviour. In order to detect real-life behaviour of interest we must observe the underlying system through a suitable set of features, and determining which features will provide the greatest separation of normal and anomalous behaviour is in itself a key challenge in anomaly detection \citep{Chandola:2009a}. This can be especially difficult in adversarial problem domains as the behaviour of anomalous entities may change over time in direct response to the detection methods employed.\\

For the detection of anomalies in social networks, we are concerned with the pattern of interactions between the individuals in the network. Thus, following \citep{Chandola:2009a,Hodge:2004} we define network anomalies simply as patterns of interaction that significantly differ from the norm. As is the case for general anomaly detection, this definition is highly imprecise, and for a given domain, the specific meaning of `pattern of interaction' and `significantly differ' will reflect the particular behaviour of interest. For example, one analysis of emails between employees of the Enron corporation considered `patterns of interaction' to simply be the number of emails sent by an individual over a given period of time \citep{Priebe:2005}. For this particular study, this metric was deemed to be a suitable feature for identifying the real-life anomalous behaviour of interest. However, for a different study, concerned with a different aspect of employee behaviour, a different metric may be considered in order to capture the appropriate `patterns of interaction'.

In this paper, we focus on the detection of anomalies resulting from patterns of interaction that differ from normal behaviour within the same social network. However, for completeness, we note here that another form of anomaly exists, termed horizontal anomalies \citep{Gao:2013}. Horizontal anomalies occur when the characteristics or behaviours of an entity vary depending on the source of the data \citep{Gao:2013,Papadimitriou:2010}. For example, a user of social media may have a similar set of friends or followers across a number of platforms (e.g. Twitter, Facebook, etc.), but for one particular platform (Google+ for example) has a markedly different set of acquaintances. Methods for detecting this type of anomaly are beyond the scope of this paper. However, given the highly diverse nature of the systems from which social network data is currently being extracted, we believe that the detection of horizontal anomalies will become extremely important in the near future, and suggest that further research in this area will be extremely fruitful.

\section{Characterisation of anomalies}
\label{NetworkAnalysis}

In analysing social networks it is the interactions between individuals that form the main object of study. Focus is given to the manner in which interactions between pairs of individuals influence interactions with and between other individuals in the system, and the relationship between these interactions and the attributes of the individuals involved (see \citet{Brandes:2013,Getoor:2005} for discussion of network analysis in general). This differentiates anomaly detection in social networks from traditional, non-network based analysis, where individual's attributes are assumed to follow some population level distribution that ignores interactions and the resulting relationships between individuals and their peers.

Typically, social networks are represented using a graph with vertices (nodes) representing individuals (or groups, companies, etc.) and edges (links, arcs, etc.) between the vertices representing interactions. In some instances, nodes may be partitioned into multiple types, and such networks can be represented using bipartite (or tripartite, etc.) graphs.\\

Depending on the type of analysis being performed, networks can be characterised as being either static or dynamic, and as being labelled or unlabelled. Obviously, dynamic networks change over time, with changes occurring in the pattern of interactions (who interacts with who), or in the nature of these interactions (how X interacts with Y). Labelled networks include information regarding various attributes of both the individuals and their interactions.


It could be argued that all social networks are dynamic, however it is often useful to analyse social networks as if they were static. As an example, consider the Enron email data set (analysed in \citet{Priebe:2005,Akoglu:2010,Eberle:2007,Eberle:2009}), which consists of a dynamic network, representing the emails sent between Enron employees between 1998 - 2002. This network changes over time as different groups of individuals send and receive emails at different times. If employee A sends five emails to employee B, a dynamic representation of the network would track these five emails as links between A and B occurring only in the appropriate time-steps. If the properties of each email are considered (e.g. length, subject, etc.) then the network is treated as being labelled. By considering a dynamic representation of the network, \citet{Priebe:2005} were able to detect anomalous time-steps where a particular individual suddenly increased the number of emails they sent relative to previous time-steps.

In contrast to a dynamic representation of the Enron network, if consideration is given to the total set of emails sent by each employee over the full time period, the time at which each email is sent is no longer deemed important and the network can be treated as static. In reducing the network from a dynamic to a static representation, multiple interactions occurring at different points in time may be combined in some way, and represented by a single link. This link may be labelled with the number of interactions it represents, and with some aggregation of the properties of these interactions (e.g. mean email length), or it may be that only the fact that an interaction took place is important, and the network can be treated as being unlabelled. For example, a static, unlabelled representation of the Enron email network was used in the analysis by \citep{Akoglu:2010}, where the formation of an anomalous star like structure was identified, indicative of a single individual (Ken Lay, a former CEO) sending emails to large numbers of individuals who were not otherwise connected (i.e. they did not send emails to one another). Thus, so long as the information is actually available, a network can be easily represented in a number of different ways, depending on the requirements of the analysis.

Exactly how a social network is chosen to be represented will depend of course on the type of anomalies to be detected. As with the networks themselves, the anomalies that occur in social networks can also be characterised as being dynamic or static, and as labelled or unlabelled. A dynamic anomaly occurs with respect to previous network behaviour, while a static anomaly occurs with respect to the remainder of the network. Labelled anomalies relate to both network structure and vertex or edge attributes, while unlabelled anomalies relate only to network structure (see \citet{Akoglu:2013b}).

In addition to the characterisation as dynamic or static and labelled or unlabelled, anomalies can be further characterised as occurring with respect to a global or local context and as occurring across a particular network unit, which we refer to as the minimal anomalous unit. A global or local context simply describes whether an anomaly occurs relative to the entire network, or relative only to its close neighbours. For example, an individual's income may be globally insignificant (many people have similar income), however if the income of their friends (defined by their links within the network) are all significantly higher, they may be considered a local anomaly \citep{Gao:2010,Ji:2012}. The minimal anomalous unit refers to the network structure that we consider to be anomalous. If we are interested in changes made by individuals, a sudden change in the medium of communication for example, we might consider a single vertex to be the minimal anomalous unit. However, if we are interested in larger groups of individuals, perhaps a group of fraudulent individuals collaborating to boost their reputation in an online auction system \citep{Pandit:2007}, the minimal anomalous unit of interest scales to that of a sub-network. In this situation, the behaviour of any given vertex may not be anomalous, but taken together as a group, the pattern of communication may be anomalous with respect to other groups of vertices in the network. Throughout this review, we will consider how the different approaches to anomaly detection relate to each of these four characteristics.

\section{Methods for anomaly detection}

In Section \ref{NetworkAnalysis}, we defined four characteristics that can be used to categorise anomalies; static or dynamic, labelled or unlabelled, local or global context, and the minimal anomalous unit. In this section we outline various approaches for detecting these different types of anomalies. In practice we found that characterisation as static or dynamic and labelled or unlabelled best differentiates the various approaches, thus the following section is organised according to characterisation along these two axes.

Note that we have considered here only methods that have previously been used to analyse social networks. Network based anomaly detection has been investigated in a number of additional problem domains including intrusion detection \citep{Garcia:2009,Gogoi:2011,Jyothsna:2011}, traffic modelling \citep{Shekhar:2001} and gene regulation \citep{Kim:2009,Kim:2012}. However since we are primarily interested in social networks we have not included these studies in our review.\\

\subsection{Static unlabelled anomalies}
\label{StaticUnlabelled}

Static, unlabelled anomalies occur when the behaviour of an individual or group of individuals leads to the formation of unusual network structures. Because the labels on edges and vertices are not considered, any information regarding the type of interaction, it's duration, the age of the individuals involved, etc. is ignored. Only the fact that the interaction occurred is significant. Thus in order to detect anomalous behaviour, assumptions must be made regarding the probability that a given pair of individuals will interact.

As an example, \citet{Shrivastava:2008} noted that email spam and viral marketing material is typically sent from a single malicious individual to many targets. Since the targets are selected at random, they are unlikely to be connected independently of the malicious individual, and the resulting network structure will form a star. Based on this observation, the number of triangles in each ego-net \footnote{An ego-net is defined as the subgraph induced by consideration of a subject and their immediate neighbours} is used to label the subject individual as being either malicious or innocent, with a low triangle count indicating a malicious individual. Extensions to this basic approach have also been used to detect groups of malicious individuals acting in collaboration \citep{Shrivastava:2008}.

While \citet{Shrivastava:2008} considered star-like structures to be indicative of malicious behaviour, \citet{Akoglu:2010} demonstrated that both near-stars and near-cliques may be indicative of anomalous behaviour in social networks. Cliques and stars form the extremal values of a power-law relationship between the number of nodes in an ego-net and the number of edges ($N_{e} \propto N_{v}^{\alpha}$), thus in order to detect anomalies, a power-law curve can be fitted to the network, and the residuals analysed for significant deviance from the expected relationship. This idea has been applied to numerous properties of the ego-net \citep{Akoglu:2010,Hassanzadeh:2012}, and in a comparison of various combinations of features, \citet{Hassanzadeh:2012} showed that a power-law relationship between edge count, and the average betweenness centrality (see \citet{Hassanzadeh:2012} for definitions) of an individual's ego-net was most useful in differentiating between normal and anomalous (near-cliques or near-stars) ego-nets.

Another approach for detecting static, unlabelled anomalies is the use of signal processing techniques. The application of signal processing stems from work on graph partitioning, and community detection \citep{Miller:2010a,Newman:2006}, and treats anomalous subgraphs as a signal embedded in a larger graph, considered to be background noise. The problem of detecting this anomalous subgraph is formulated in terms of a hypothesis test \citep{Miller:2010a,Miller:2010b,Miller:2011a} with

\begin{quote}
H0: \textit{the graph is noise (no anomalous subgraph exists)}\\
H1: \textit{the graph is signal + noise (an anomalous subgraph is embedded in the graph)}
\end{quote}

\noindent The null hypothesis assumes that the network is generated by some stochastic process describing the probability that any pair of nodes will be connected by an edge. Anomalous subgraphs are generated by different underlying processes, resulting in a density of edges that differs from that expected under the null model \citep{Miller:2010a}. Clearly, the degree to which a signal processing approach can be successfully applied depends on the identification of suitable null models describing the background graph. The most basic model of random graphs, the Erd\"{o}s-R\'{e}nyi model, treats the possibility of an edge between two vertices as an independent Bernoulli trial, so that any given edge exists with probability $p$. However, graphs generated using this process exhibit a Poisson distribution of vertex degrees, which has been shown to be a poor approximation of many real world networks. Typically, real social networks exhibit a much fatter tail than the Poisson distribution \citep{Newman:2001,Menges:2008}. More sophisticated models, such as the recursive-matrix model (R-MAT) are able to generate more realistic graphs with some degree of clustering and non-Poisson degree distributions \citep{Chakrabarti:2004,Newman:2001,Menges:2008}. However, the selection of a suitable model remains a challenge for signal processing approaches, and further research is required in this area.

\subsection{Static labelled anomalies}

By considering vertex and edge labels in addition to the network structure, a context can be defined in which typical structures may be considered anomalous. Conversely, the particular combination of edge and vertex labels within the context defined by a given network structure (say an ego-net or community) may also be considered anomalous.

In addition to the detection of cliques and stars using unlabelled properties, \citet{Akoglu:2010} also considered edge labels to detect `heavy' ego-nets, where the sum over a particular label is disproportionately high relative to the number of edges. For example, in analysing donations to US presidential candidates, the Democratic National Committee were found to have donated a substantial amount of money, but this money was spread over only a few candidates, thus the ego-net formed is considered `heavy'. In contrast the Kerry Committee received a large number of donations, but each donation was for a small amount. In this example, a bipartite network is considered (with candidates and donors as the two vertex types), and the ego-net provides a context in which the received or donated amounts may be considered anomalous. Outside this structure, a series of small or large donations may be perfectly normal.

Taking a similar view of context, the detection of community dependent anomalies has been investigated by dividing the network into communities based on individuals' patterns of interactions, and then considering the attributes of the members of each community \citep{Gao:2010,Ji:2012,Gupta:2012,Muller:2013}. Attribute values which would be deemed normal across the entire network may appear as anomalies within the more restricted context imposed by the community setting. For example, \citet{Gao:2010} describe a situation where the patterns of interaction between a particular group of individuals marks them as a community within a larger network, however one member of this community has a far lower income than their peers. Globally, this individual's income is completely normal, however within the context provided by the community their income is seen as an anomaly.\\

In the realm of spam detection, static labelled anomalies have been used to identify opinion spam in online product reviews \citep{Akoglu:2013,Chau:2006,Pandit:2007}. These researchers have applied belief propagation to anomaly detection, whereby `hidden' labels assigned to vertices and edges are updated in an iterative manner conditionally based on the labels observed in the system of interest.

As an example of belief propagation, the FraudEagle algorithm \citep{Akoglu:2013} uses a bipartite graph to represent reviews of products in an online retail store, with users forming one set of vertices, and products forming the other. Edges between users and products represent product reviews and are signed according to the overall rating given in the review. The aim of FraudEagle is to assign a hidden label to each node, with users labelled as either honest or fraudulent, and products as good or bad. These labels are assigned based on the observed pattern of reviews. It is assumed that normal, honest reviewers will typically give good products positive reviews and bad products negative reviews, while fraudulent reviewers are assumed to regularly do the opposite. This assumption is encoded as a matrix describing the probability of a user with a hidden label of either honest or fraudulent giving a positive or negative review of a product whose hidden label designates it as either good or bad. By iteratively propagating labels through the network according to these assumed probabilities (i.e. by applying loopy belief propagation), the system is able to determine those users who can be considered as opinion spammers.

While not strictly applied to social networks, TrustRank \citep{Gyongyi:2004} could also be considered as an example of a belief propagation approach to anomaly detection. In TrustRank, trustworthy pages are assumed to be unlikely to link to spam pages, and given an initially labelled set of trustworthy pages, trust is propagated through page out-links, so that those pages within k-steps of the the initially trusted page are also labelled as trustworthy. By ranking pages according to the propagated trust score, spam pages, representing anomalies, can be detected.\\

Another approach to the detection of static labelled anomalies is the application of information theory, which provides a set of measures for quantifying the amount of information described by some event or observation. The most commonly used measure, entropy, describes the number of bits required to encode the information associated with a given event. Thus, given a set of observations O, the global entropy H(O) can be thought of as a measure of the degree of randomness or heterogeneity. A set containing a large number of different observations will require more bits to store the associated information and will therefore have a higher global entropy. In contrast a totally homogeneous set will require relatively few bits to store the associated information.

In terms of anomaly detection, entropy can be used to identify those subgraphs that if removed from the network would lead to a significant change in entropy. These subgraphs may represent important communities or individuals \citep{Shetty:2005,Tsugawa:2010,Serin:2012}, or may represent a normal pattern of behaviour that can then be used as a point of comparison, with those subgraphs that deviate from this norm considered to be anomalous \citep{Noble:2003,Eberle:2007,Eberle:2009}. In comparing subgraphs, an information theoretic approach considers a global context, using a very different representation of the network than that used in the other approaches described in this paper. In applying an information theoretic approach, interactions are represented as subgraphs, and values typically represented as labels on edges or vertices are themselves represented as a vertex in this subgraph.

Information theoretic approaches have been shown to be capable of detecting anomalous interactions based on complex structures and flows of information. For example, a version of this method was applied to the Enron email data set \citep{Eberle:2009}, resulting in detection of a single instance of an email being sent from a director to a non-management employee, and then being forwarded to another non-management employee. Within the Enron data set many emails are sent from directors to non-management employees and between pairs of non-management employees, but this was the only instance involving a forwarded email from a director to a non-management employee. This degree of sensitivity is useful in problem domains consisting of adversarial systems where individuals actively seek to hide illegal or otherwise unwanted behaviour, and deviations from the norm are likely to be small \citep{Eberle:2009}.

\subsection{Dynamic unlabelled anomalies}

Dynamic unlabelled anomalies arise when patterns of interaction change over time, such that the structure of a network in one time-step differs markedly from that in previous time-steps. As with static unlabelled anomalies, there are obviously numerous graph properties that we may consider, each of which can be used to generate time-series that can be analysed using traditional tools. Indeed, we could go so far as to use a static analysis (e.g. the parameter of a fitted power-law relationship) as the subject of a time-series and consider how this value changes over time. The difficulty lies, of course, in selecting an appropriate feature, or set of features, that adequately captures the real-world behaviour of interest.

Within the literature we found examples of dynamic unlabelled network analysis using a variety of methods, including scan statistics \citep{Priebe:2005,Neil:2011,Cheng:2013,Marchette:2012,Park:2009,McCulloh:2011}, Bayesian inference \citep{Heard:2010}, auto-regressive moving average (ARMA) models \citep{Pincombe:2005}, and link prediction \citep{Huang:2006}. The use of these methods has mostly focused on individual nodes, or ego-nets, considering the node degree or the size of the ego-net.

Link prediction attempts to predict future interactions between individuals based on previous interactions. In order to detect anomalies, a link prediction algorithm is run, and for each pair of individuals in the network, the likelihood of interaction occurring in the next time-step is calculated. This predicted likelihood is then compared to the observed interactions, and those observed interactions having a low predicted likelihood are deemed anomalous \citep{Huang:2006}.

Using scan statistics, analysis is performed by averaging a time-series over a moving window $[t - \tau, t]$ and comparing this average to the current time-step. If the difference is larger than some predefined threshold, the time-step is considered to be anomalous \citep{Priebe:2005,Neil:2011,Cheng:2013,Marchette:2012,Park:2009}. The use of ARMA models involves fitting the model to a time series and considering the residuals for this fit (see \citet{Pincombe:2005}). Those residuals above a specified cut-off are deemed to represent anomalous time-steps.

In Bayesian analysis, the generated time series is assumed to follow some distribution (i.e. each time-step represents a draw from the assumed distribution). Given an initial estimation of the distribution parameters, a series of updates can be performed in an iterative manner. For each time step, the prior distribution is updated by considering the value at that time step and calculating a posterior distribution that incorporates the information from all of the previous time steps. Using this posterior distribution, a p-value is calculated, giving the probability that the value at the next time step was drawn from this distribution. If this p-value is below a significance threshold, the value can be considered anomalous. The posterior distribution is then treated as the prior for the next update \citep{Heard:2010}.\\

The use of scan statistics is the most prominent form of analysis we found in the literature, and extensions to the basic form have been suggested that allow correlated changes in graph properties to be detected \citep{Cheng:2013}. This enables inter-related vertices or subgraphs to be identified, and may be of benefit in detecting clandestine organisations. Another extension is the application to hypergraphs \citep{Park:2009}, which are a generalisation of graphs allowing hyperedges that can connect more than two vertices. Such graphs can be used to represent interactions that involve more than two individuals such as meetings, group emails and text messages, and co-authorship on research papers \citep{Park:2009,Silva:2008}.

Other researchers have considered larger structures such as maximal cliques \citep{Chen:2012}, and coronets \footnote{A coronet is a concept of neighbourhood, where the `closeness' to a subject is defined in terms of the number of interactions that occur between individuals, rather than simple connectedness. See \citep{Ji:2013} for a formal definition. Note that the number of interactions between two individuals is often expressed as a weight, which could be considered a label on the edge. However, since multiple interactions can also be represented by multiple edges, we do not consider this to be a true label.} \citep{Ji:2013}, with a view to detecting anomalous evolution of a system at a neighbourhood scale. If we consider only the pattern of interactions between individuals, a maximal clique can evolve in only six different ways; by growing, shrinking, merging, splitting, appearing or vanishing \citep{Chen:2012}. Similarly, a coronet can evolve only through the insertion or deletion of a node, or the insertion or deletion of a number of edges. If normal behaviour is assumed to result in stable, non-evolving neighbourhoods, then any neighbourhood that undergoes one of these transformations can be considered to be anomalous. For less stable systems, this idea could obviously be extended to include consideration of the number of neighbourhoods that change over a time-step, or the magnitude of this change. In such scenarios, scan-statistics, Bayesian inference or any other form of time-series analysis may be employed in order to detect anomalous time-steps.\\

In addition to the approaches discussed above, there is a large body of work that focuses on anomaly detection in time-series in general (see \citet{Cheboli:2010,Chandola:2009b} for recent reviews). We believe that these existing methods could also be used to detect dynamic anomalies in social networks, assuming that suitable features can be identified for generating the required time-series. Moreover, substantial work has been undertaken in the field of change detection in social networks (e.g. \citet{McCulloh:2011,Arney:2013,Tambayong:2014}, and we believe that there is considerable room for cross-over between these two fields.

\subsection{Dynamic labelled anomalies}

In conducting our survey, we found dynamic labelled anomalies to be the least represented in the literature. A recent paper describes the application of a signal processing approach to anomaly detection in dynamic, labelled networks \citep{Miller:2013}, however this was the only example we found.

Extending the basic signal processing approach used for static, unlabelled anomalies \citep{Miller:2010a,Miller:2010b}, detection of dynamic labelled anomalies assumes that the probability of an edge occurring between any two nodes is a function of the the linear combination of node attributes \citep{Miller:2013}. Coefficients for this linear combination are fitted to the network of interest. The dynamic nature of the network is handled by considering the network structure in discrete time-steps, and treating each time-step as for a static network. Comparing the expected graph to the observed graph in each time-step gives a residual matrix that can be integrated over a specified time period using a filter (see \citet{Miller:2013,Miller:2011b} for details). Note that this method assumes the number of vertices remains constant over the specified time-period, and that only the edges change.\\

In addition to the signal processing approach described above, we suggest that those methods used for detecting dynamic unlabelled anomalies could be adapted for the detection of dynamic labelled anomalies. We argue that this type of development will be extremely beneficial as including additional information provided by node and edge labels is likely to significantly improve anomaly detection, providing additional dimensions for separating normal and anomalous behaviour. For example, a recent study demonstrated that link prediction can be significantly improved by including topic information from users tweets \citep{Rowe:2012}. Obviously if the algorithm used to predict links is improved, any anomaly detection using this algorithm will also be improved.

As with dynamic unlabelled anomalies, we also suggest that existing methods for time-series analysis can be applied to the detection of dynamic labelled anomalies. A number of methods for detecting anomalies within discrete sequences are discussed in \citep{Chandola:2012}, and we believe that many of these methods could be applied to the detection of dynamic labelled anomalies in social networks.

\section{Discussion}
\label{Discussion}

In this paper we have surveyed a small but growing number of solutions to detecting anomalies in online social networks. While these approaches differ in their treatment of the network, considering dynamic or static representations and including or ignoring node attributes, they are all based primarily on the analysis of interactions between individuals. It is this basis that differentiates the detection of anomalies in social networks form other forms of anomaly detection.

Note that while anomalies and networks can be characterised as static or dynamic, labelled or unlabelled etc., the various approaches to anomaly detection can, of course, also be grouped by the types of algorithms employed and also the problem domain they were originally developed to address. In conducting our survey, we found it quite interesting that the partitioning of detection methods by algorithm type very nearly matches the partition induced by the authorship of related literature. During our review, we found very little crossover between the different research groups exploring this area, with each group tending to focus on a single or limited number of approaches. Certainly, we did not find clear lines of development, where a particular approach has evolved through incremental improvements over a large number of iterations. Clearly, this reflects the young nature of this field, and we anticipate that this state of affairs will change significantly over the next decade.\\

In considering how the various approaches presented in the literature apply to different types of network anomalies, we found that the methods used to actually detect anomalies, as opposed to calculating network features, is largely independent of the type of anomaly being detected. Once a suitable feature space has been identified and calculated, it is often the case that any number of traditional anomaly detection techniques could be applied. For example, in Section \ref{StaticUnlabelled} (static unlabelled anomalies), we discuss the use of regression to detect anomalous ego-nets displaying an unusual edge count relative to node count \citep{Akoglu:2010}. However, rather than applying regression, once the edge count - node count feature space is extracted, an alternative anomaly detection method, such as a nearest neighbour method, could just as easily have been applied. Therefore, we argue that the process of detecting anomalies in social networks is composed of two quite separable sub-processes; namely, the calculation of a suitable feature space, and the detection of anomalies within this space. Moreover, we argue that the selection of a suitable feature space can be further broken down so that, taking a high level view of the problem, the vast majority of the approaches we reviewed can be generalised into the following five steps.

\begin{enumerate}
\item Determine the smallest unit affected by the behaviour of interest (node, neighbourhood, community, etc).
\item Identify the particular properties of this unit that are expected to deviate from the norm.
\item Identify the context in which these deviations are expected to occur.
\item Calculate the properties of interest, extracting a feature space in which the distances between observations can be measured and compared in a meaningful way.
\item Within this space, calculate distances between observations, possibly applying traditional (non-network) tools for anomaly detection (e.g. clustering, nearest neighbour analysis etc.).
\end{enumerate}

In each of the papers we surveyed we found something akin to the five steps given above. Obviously there is a major focus on step 4, as the naive approach to this step is often computationally expensive. Given the size of many online social networks, and the demand for tools to handle Big Data we see the development of scalable anomaly detection solutions as being an extremely important area of research. However, we also see the selection of a suitable feature space (steps 1 - 3) as being equally important.

We believe that our steps 1 - 3, which encompass the mapping of real-world behaviours of interest to suitable feature spaces, form a key challenge for detecting anomalies in social networks. In conducting our survey, few of the papers we reviewed explicitly described how the mapping between the behaviour of interest and the network properties considered was performed. Those few papers that did (e.g. \citet{Akoglu:2013,Shrivastava:2008,Friedland:2009,Gao:2010,Ji:2013}), clearly explained the combination of domain knowledge and mathematical reasoning that underpin the use of a particular approach and the examination particular network features. For the majority of papers though, the reasons for considering a particular set of features were unclear, and in many examples the real-world behaviour giving rise to any anomalies detected was discussed only in response to the anomaly actually being found, rather than as a motivation for using a particular approach. We believe that this is problematic as many of the approaches we reviewed were tested only on a single, or a limited number of data sets, and may therefore be biased towards the particular anomalies inherent in those data sets. Without explicit reasoning as to why a particular behaviour should be identifiable by a particular anomalous network feature there is no reason to suspect that this feature will behave in a particular manner across different data sets, representing different social networks.

The lack of papers clearly describing the reasons for examining a particular set of features suggests to us that selection of a suitable feature space may be extremely difficult in practice. Many properties of social networks are correlated in some fashion, and it is not clear which combinations of features can be used to capture orthogonal concepts. Some comparison of feature spaces has been undertaken (e.g. \citet{Hassanzadeh:2012,McCulloh:2011}), however such comparisons have been restricted in the number of features considered, and the types of anomalies being detected. We predict that the requirements for anomaly detection in social networks will rapidly advance in the near future, with ever larger volumes of data and increasingly complex behaviours being considered. This may in turn lead to the consideration of increasingly complex feature spaces. Thus the development of any guidelines or heuristics for mapping real-world behaviour to appropriate feature spaces will be of benefit. One possibility that we see for identifying suitable feature spaces is the application of a shotgun approach. By applying multiple detection techniques across numerous graph properties, it may be possible to identify mechanistic links between the actual behaviours of interest and resulting changes in network properties, and to determine those features that best differentiate anomalous and normal behaviours.

Another major challenge for detecting anomalies in online social networks is the evaluation and comparison of different methods. Clearly, not all algorithms are universally applicable, and many of the existing approaches have been developed with specific problem domains and data formats in mind. Moreover, there is a distinct lack of publicly available data sets with known ground truths. In order to compare existing and novel techniques, it is important to consider how accuracy (recall) and precision are influenced by the scale of the analysis, the density of anomalies, and the magnitude of difference between anomalous and normal observations. However, the range of different data sets required to perform such comparisons is not currently available. Thus many approaches are tested on a single data set, with verification of results performed for the top ranked anomalies using manual investigation. For example, \citet{Pandit:2007} analyse transactions in an online auction system, and in order to test their approach results were ranked according to an anomaly score. The top-ranked user accounts were then manually investigated by checking the auction website for complaints made against these accounts. This type of manual investigation is highly time-consuming and highly dependent on the level of reporting by the owners of the target system. In many cases, manual investigation of user accounts in this way is impossible. 

Some data sets do exist that are relatively well characterised, including the Enron email data set and to a lesser extent the Paraiso communication (synthetic, VAST 2008, www.cs.umd.edu/hcil/VASTchallenge08/). Anomalies detected within these data sets can be easily verified against the known series of events. However these data sets are relatively small, and may only be suitable for a small subset of problem domains. Therefore, in order to address this challenge, we believe that the generation of synthetic data sets is a logical course of action.

A number of methods for generating random social networks have been proposed, ranging from simple models assuming some global probability that a given pair of nodes will be connected to more complex models considering the dynamics of each individual node \cite{Menges:2008}. Between these two extremes lies a broad spectrum of models that have been developed over time and thus reflect our growing understanding of the general properties of social networks (e.g. high transitivity, positive, assortivity, presence of community structures). Consequently, early models (reviewed by \citet{Chakrabarti:2006}, see also \citet{Sallaberry:2013}) capture only a small number of what we now consider to be general properties of social networks \cite{Akoglu:2009}, while more recent models (e.g. \cite{Akoglu:2009,Sallaberry:2013}) are able to reproduce far more realistic representations of a wide range of social networks. As an example, \citet{Akoglu:2009} describe eleven properties of social networks which they suggest should be reproduced by random network generators and provide a suitable method for generation that does indeed reproduce these properties. This method is reported to be highly flexible, and can be used to generate weighted or unweighted, directed or undirected and unipartite or bipartite networks. The method has since been applied in the testing of a novel anomaly detection method, being used to generate a bipartite network representing user reviews of online products in an e-commerce setting \citep{Akoglu:2013} (a method for generating these types of networks is also described as part of the Berlin SPARQL Benchmark \url{http://wifo5-03.informatik.uni-mannheim.de/bizer/berlinsparqlbenchmark}, last accessed May, 2014).

The vast majority of network generation methods proposed to date can be described as phenomenological in nature. While these methods are able to reproduce many of the known properties of social networks, they do so without describing the underlying mechanisms that cause these properties to arise. In contrast, agent-based models (alternatively, individual-based models) such as that proposed by \citet{Menges:2008} (see also \citet{Glasser:2013,Hummel:2012}) describe the specific actions taken by individuals in the underlying system that lead to the formation of a social network. In this type of simulation, the macro-level properties of the resulting social network can be seen as emergent properties stemming from the micro-level behaviour of the underlying agent-based system. The model does not explicitly set out to reproduce these properties, but rather to represent the behaviour of the individuals involved. In the course of representing this behaviour, the model reproduces a realistic social network. We believe that this approach has great potential for generating realistic data sets and represents an extremely interesting area of research, particularly in adversarial problem domains where anomalous agents could attempt to disguise their behaviour by mimicking 'normal' agents.\\

In this paper we have surveyed a number of approaches for detecting anomalies in online social networks. We found that these different approaches can be usefully categorised based on characterisation of anomalies as being static or dynamic and labelled or unlabelled. Depending on this characterisation, different features of the network may be examined, and we have suggested that the selection of appropriate features is a highly non-trivial task. We believe that as the requirements of social network analysis continue to grow, the challenges inherent in Big Data analysis will necessitate sophisticated heuristics and algorithms for mapping complex behaviours to network features, and scalable solutions for calculating rich feature spaces which can be subjected to anomaly detection analysis.

\section{References}

\bibliography{refs}

\end{document}